\documentclass[12pt,preprint]{aastex}

 \usepackage{epstopdf}

 \slugcomment{Accepted Version 1}

\newcommand\pv{\mbox{$p_V$}}
\newcommand\CV{\mbox{$V_{p}$}}
\newcommand\CO{\mbox{$O_{p}$}}
\newcommand\CQ{\mbox{$Q_{p}$}}
\newcommand\CS{\mbox{$S_{p}$}}
\newcommand\CA{\mbox{$A_{p}$}}
\newcommand\CL{\mbox{${\cal{L}}_{p}$}}
\newcommand\CD{\mbox{$D_{p}$}}
\newcommand\CX{\mbox{$X_{p}$}}
\newcommand\CC{\mbox{$C_{p}$}}
\newcommand\irfactor{\mbox{$p_{IR}/p_{V}$}}
\newcommand\boznemcova{\mbox{Bo\v{z}n\v{e}mcov\'{a}}}

\begin{document}

 \DeclareGraphicsExtensions{.pdf,.gif,.jpg}

 \title{NEOWISE Studies of Asteroids with Sloan Photometry: Preliminary Results}
\author{A. Mainzer\altaffilmark{1}, J. Masiero\altaffilmark{1}, T. Grav\altaffilmark{2}, J. Bauer\altaffilmark{1}$^{,}$\altaffilmark{3}, D. J. Tholen\altaffilmark{4}, R. S. McMillan\altaffilmark{5}, E. Wright\altaffilmark{6}, T. Spahr\altaffilmark{7}, R. M. Cutri\altaffilmark{3}, R. Walker\altaffilmark{8}, W. Mo\altaffilmark{2}, J. Watkins\altaffilmark{6}, E. Hand\altaffilmark{1}, C. Maleszewski\altaffilmark{5}}

\altaffiltext{1}{Jet Propulsion Laboratory, California Institute of Technology, Pasadena, CA 91109 USA}
\altaffiltext{2}{Department of Physics and Astronomy, Johns Hopkins University, Baltimore, MD 21218 USA}
\altaffiltext{3}{Infrared Processing and Analysis Center, California Institute of Technology, Pasadena, CA 91125 USA}
\altaffiltext{4}{Institute for Astronomy, University of Hawaii, 2680 Woodlawn Drive, Honolulu, Hawaii 96822-1839 USA}
\altaffiltext{5}{Lunar and Planetary Laboratory, University of Arizona, 1629 East University Blvd., Kuiper Space Science Bldg. \#92, Tucson, AZ 85721-0092 USA}
\altaffiltext{6}{UCLA Astronomy, PO Box 91547, Los Angeles, CA 90095-1547 USA}
\altaffiltext{7}{Minor Planet Center, Harvard-Smithsonian Center for Astrophysics, 60 Garden Street, Cambridge, MA 02138 USA}
\altaffiltext{8}{Monterey Institute for Research in Astronomy, Monterey, CA USA}

 \begin{abstract}
 We have combined the NEOWISE and Sloan Digital Sky Survey data to study the albedos of 24,353 asteroids with candidate taxonomic classifications derived using Sloan photometry.  We find a wide range of moderate to high albedos for candidate S-type asteroids that are analogous to the S-complex defined by previous spectrophotometrically-based taxonomic systems. The candidate C-type asteroids, while generally very dark, have a tail of higher albedos that overlaps the S types.  The albedo distribution for asteroids with a photometrically derived Q classification is extremely similar to those of the S types.  Asteroids with similar colors to (4) Vesta have higher albedos than the S types, and most have orbital elements similar to known Vesta family members.  Finally, we show that the relative reflectance at 3.4 and 4.6 $\mu$m is higher for D-type asteroids and suggest that their red visible and near-infrared spectral slope extends out to these wavelengths.  Understanding the relationship between size, albedo, and taxonomic classification is complicated by the fact that the objects with classifications were selected from the visible/near-infrared Sloan Moving Object Catalog, which is biased against fainter asteroids, including those with lower albedos.   
 
 \end{abstract}

 \section{Introduction}
 Asteroids and comets represent the leftovers from the formation of our Solar System.   By studying their compositional variation, we can begin to better understand the conditions present at the earliest stages of planet formation as well as their subsequent evolution and processing.  Asteroids are grouped into three main categories: C type or carbonaceous asteroids, thought to be the most common type in the Main Belt; the S type or stony asteroids, a spectrally diverse group, and the X types, another diverse group of asteroids that have relatively featureless spectra but a wide range of albedos, probably representing a broad range of mineralogies and thermal histories.  \citet{Gaffey93M} give an overview of some of the earlier taxonomic systems \citep[e.g.][]{Chapman75,Bowell78,Barucci,Tedesco89a,Tedesco89b,Howell}.  \citet{Tholen84} created 14 taxonomic classes based on the combinations of colors available from the Eight-Color Asteroid Survey \citep[ECAS;][]{Zellner,Tholen89}. ECAS used filter passbands with wavelengths ranging from 0.34 to 1.04 $\mu$m, since UV wavelengths were detectable by the photomultiplier available at the time.  The Tholen and Tedesco systems also used albedo to differentiate asteroids; the X-group of asteroids have similar ECAS colors but markedly different albedos.  
 
The Tholen and Tedesco systems offer a powerful means of distinguishing unique asteroid groups.  However, when CCDs replaced the photomultipliers at most observatories, they ushered in a factor of several improvement in quantum efficiency at red wavelengths compared to blue; while the ECAS photomultipliers extended out to 1.1 $\mu$m, most CCDs' responses were diminished at this wavelength.  These effects combined to make it difficult for observers to collect the full set of UV and near-infrared bands used by ECAS.  Groups such as \citet{Xu}, \citet{Lazzaro} and \citet{Bus02a} obtained visible spectra of asteroids and used them for classification without albedos.  \citet{Bus02b} created a new taxonomic scheme relying solely on visible spectroscopy with 26 taxonomic types, and \citet{DeMeo09} revised that scheme to 24 taxonomic types based on visible and near-infrared (VNIR) spectral signatures.  Neither the \citet{Bus02b} nor \citet{DeMeo09} systems rely upon having albedo measurements for classification.  Nevertheless, systems based solely on VNIR spectroscopy and photometry are widely used because observations at these wavelengths have generally been far more widely available than albedo measurements for the $\sim$500,000 asteroids known today. However, as discussed in \citet{Gaffey02}, the ability to link taxonomic classification to asteroid mineralogy is complicated by the unknown surface properties such as particle size, the fact that some minerals have limited or no spectral features in the wavelengths used for classification, overlapping features, etc.  It is therefore useful to try to understand how the various taxonomic types are linked to physical properties such as albedo and density.   
  
 The fourth release of the Sloan Digital Sky Survey (SDSS) Moving Object Catalog \citep[MOC;][]{Stoughton,Abazajian} provided near-simultaneous observations of $\sim$100,000 known asteroids in five bands ($u$, $g$, $r$, $i$ and $z$) \citep{Ivezic}, and these have been used to study the distribution of colors throughout the Main Belt \citep[c.f.][]{Parker,Nesvorny}.  These studies have been conducted largely without reference to albedo, simply because the number of asteroids in the SDSS MOC vastly outnumbers those with well-measured albedos.  To date, the \emph{Infrared Astronomical Satellite} \citep[IRAS;][]{Tedesco02} has been the largest source of radiometrically measured asteroid albedos, providing measurements for $\sim$2000 asteroids.   

With the \emph{Wide-field Infrared Survey Explorer's} NEOWISE project \citep{Wright, Mainzer11a}, thermal observations of $>$157,000 asteroids  are now in hand.  In \citet{Mainzer11d}, we compared the albedos derived from NEOWISE observations of $\sim$1900 asteroids with taxonomic types derived from VNIR spectroscopy. Here, we use the SDSS MOC photometry to obtain approximate taxonomic classifications for asteroids with NEOWISE observations.  We initially focus on the system of \citet{Carvano}, who define a classification algorithm based on SDSS colors that is compatible with previous taxonomic systems.  

\section{Observations}
WISE surveyed the entire sky in four infrared wavelengths, 3.4, 4.6, 12 and 22 $\mu$m (denoted $W1$, $W2$, $W3$, and $W4$ respectively).  Descriptions of the pre-launch mission design and testing can be found in \citet{Liu, Mainzer}, and the post-launch description is given in \citet{Wright}.  A series of enhancements to the WISE data processing pipeline, known as NEOWISE, have enabled the detection of $>$157,000 asteroids and comets throughout the Solar System, including the discovery of $\sim$34,000 new minor planets \citep{Mainzer11a}. 

24,353 objects, including nine NEOs and $\sim$24,275 Main Belt asteroids (MBAs), were detected during the fully cryogenic portion of the NEOWISE survey and had matches with SDSS MOC observations of sufficient quality to enable classification according to the method described in \citet{Carvano}.  The observations of these objects were extracted from the WISE archive using the First Pass version of the WISE data processing pipeline \citep{Cutri} following the methods and parameters given in \citet[][hereafter M2]{Mainzer11b}, \citet[][M3]{Mainzer11c}, and \citet[][M4]{Mainzer11d}.  

\section{Preliminary Thermal Modeling}
We have created preliminary thermal models for each asteroid using the First-Pass Data Processing Pipeline described above.  As described in references M2, M3 and M4, we employ the near-Earth asteroid thermal model (NEATM) of \citet{Harris}.  The NEATM model uses the so-called beaming parameter $\eta$ to account for cases intermediate between zero thermal inertia \citet[the Standard Thermal Model; ][]{Lebofsky_Spencer} and high thermal inertia \citet[the Fast Rotating Model; ][]{Lebofsky78,Veeder89,Lebofsky_Spencer}. With NEATM, $\eta$ is a free parameter that can be fit when two or more infrared bands are available. Bands $W1$ and $W2$ typically contain a mix of reflected sunlight and thermal emission. The flux from reflected sunlight was computed for each WISE band using the methods described in M2, M3 and M4; when sufficient reflected sunlight was present in bands $W1$ and $W2$, it was possible to compute the reflectivity at these wavelengths, $p_{IR}$, where we make the assumption that $p_{IR}=p_{3.4}=p_{4.6}$.  The validity of this assumption and the meaning of $p_{IR}$ is discussed in M4 and will be the subject of future work. As described in M2 and M3, the minimum diameter error that can be achieved using WISE observations is $\sim10\%$, and the minimum relative albedo error is $\sim20\%$ for objects with more than one WISE thermal band for which $\eta$ can be fitted.  For objects with large amplitude lightcurves, poor $H$ measurements, or poor signal to noise measurements in the WISE bands, the errors will be higher.
 
\subsection{High Albedo Objects}
We note that among the 24,353 asteroids considered here, there are $\sim$55 that have $p_{V}>0.65$. Of these, 48 have $W3$ peak-to-peak variations $>$0.3 mag, indicating that they are likely to be highly elongated. Almost all of the extremely high albedo objects have orbital elements consistent with membership in either the Vesta family or the Hungarias.   \citet{Harris1988} and \citet{Harris1989} noted that E and V type asteroids can have slope values as high as $G\sim$0.5.  The assumption that we have used of $G=0.15$ for an object like this would cause an error in the computed $H$ for observations at 20$^{\circ}$ phase angle of $\sim$0.3 magnitudes; this would drive the albedo derived using such an $H$ value up by 0.3.   These objects would greatly benefit from an improved determination of their $H$ and $G$ values.   

\section{Discussion}
The Carvano scheme uses the SDSS colors as well as their measurement uncertainties to define nine spectral classes: \CV, \CO, \CS, \CA, \CL, \CD, \CX, \CQ and \CC.  These are roughly analogous to the spectroscopically defined systems of \citet{Bus02b} and \citet{DeMeo09}; the ``p'' indicates that the classification was derived photometrically.   The nine Carvano classes were defined based on the ability of the SDSS colors to represent the system of \citet{Bus02b}.  The exception to this is the \CL\ class, which is a conglomeration of the Bus K, L and Ld classes.  For each SDSS observation, the probability that an object could be associated with a particular class was computed using the five SDSS magnitudes and their associated uncertainties; classifications and probabilities for $\sim$63,000 asteroids were taken from \citet{Hasselmann}.  

\begin{figure}
\figurenum{1}
\includegraphics[width=6in]{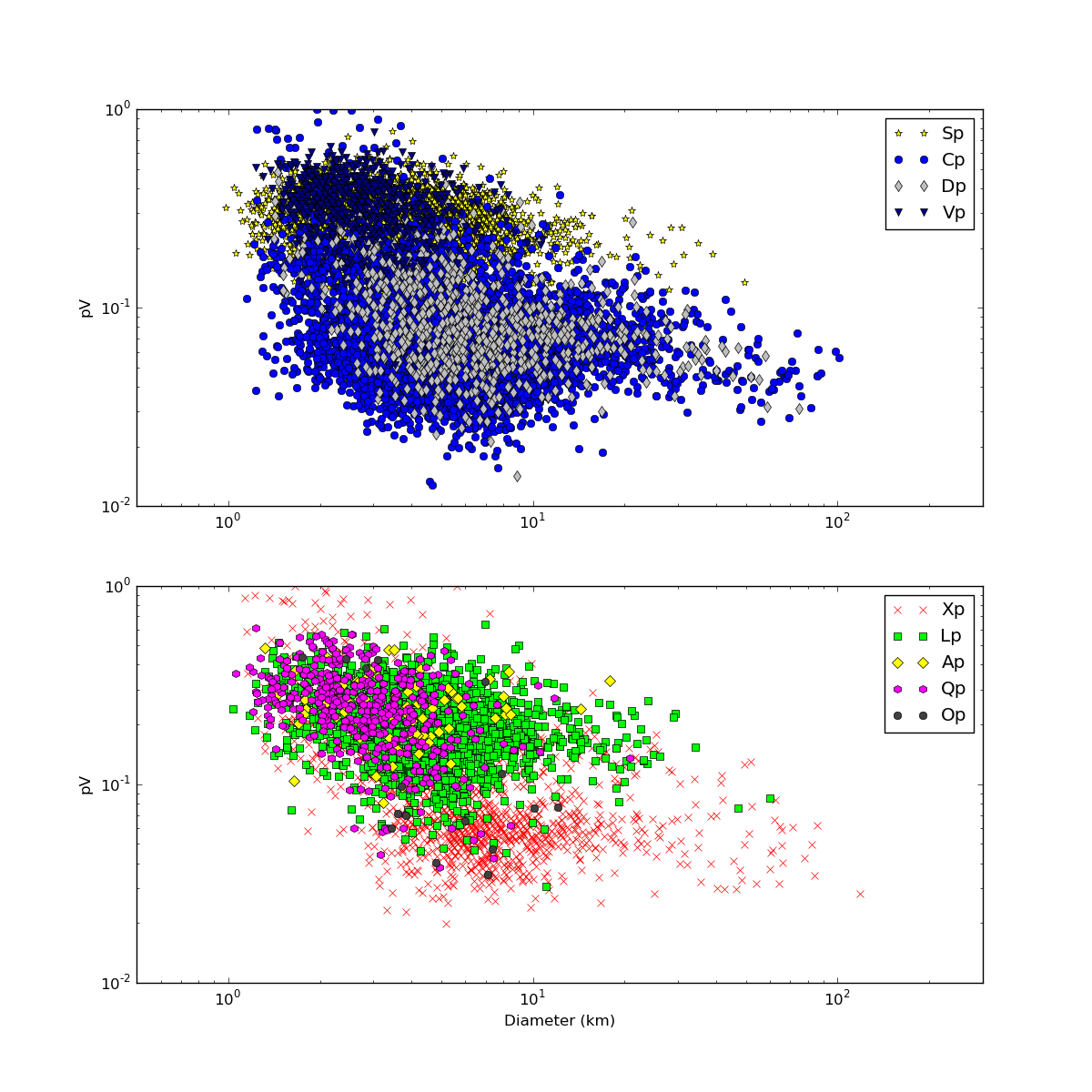}
\caption{\label{fig:diam_alb} Visible albedo as a function of size for the nine individual classes defined by \citet{Carvano}.  The classes have been separated into two sub-panels for clarity.  The bias of the SDSS survey against small, low albedo objects is evident. }
\end{figure}

\begin{figure}
\figurenum{2}
\includegraphics[width=6in]{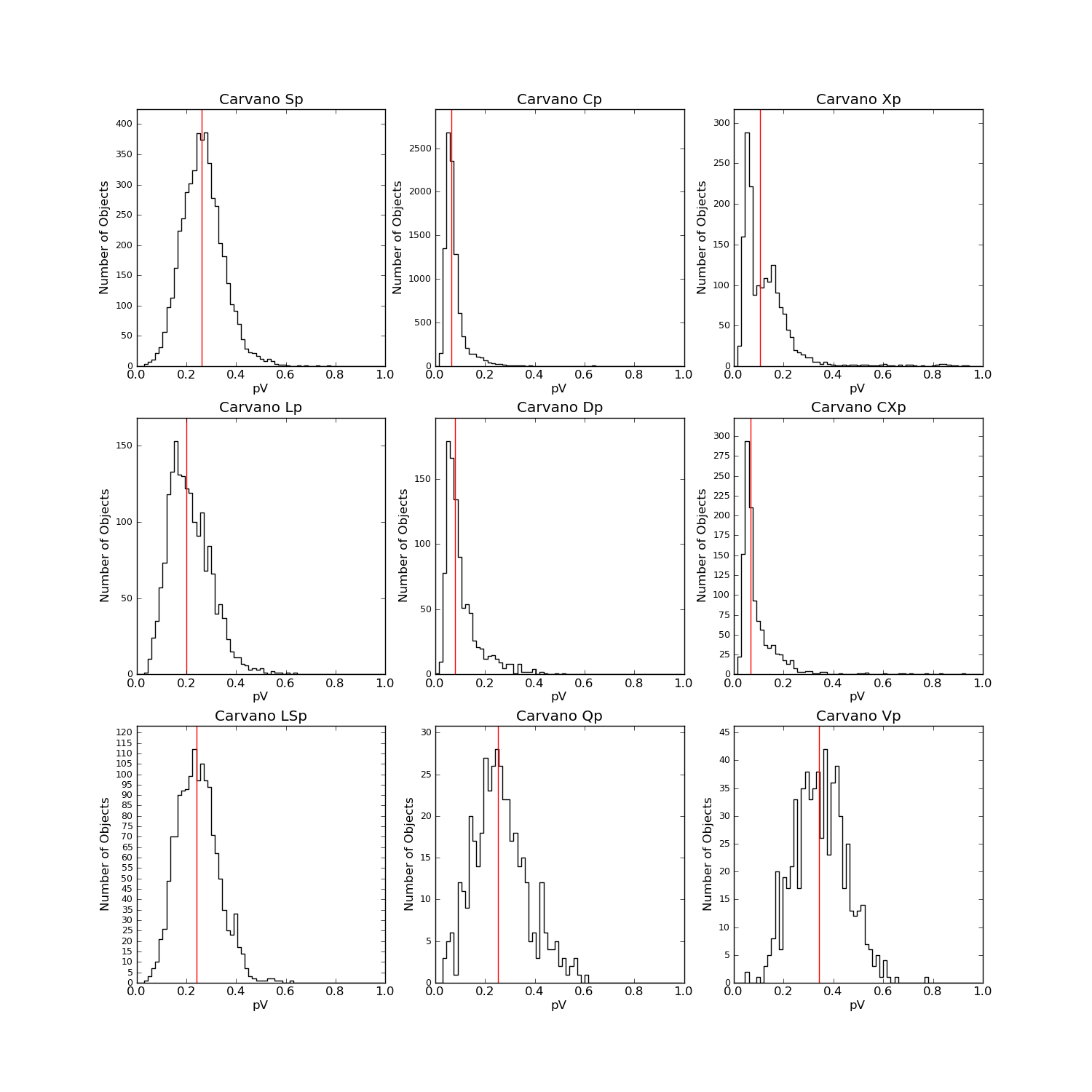}
\caption{\label{fig:carvano_albedo} NEOWISE-derived albedos of asteroids observed and classified by \citet{Carvano}.  Only classes with more than $\sim$100 asteroids are plotted, and these include some objects that have multiple SDSS observations that produced different classifications. The median \pv\ value is shown as a vertical red line.}
\end{figure}

Figure \ref{fig:diam_alb} shows \pv\ as a function of diameter for the individual Carvano classes. The bias of the VNIR SDSS survey against small, low albedo asteroids is evident in the dearth of objects in this regime, despite the fact that low albedo objects dominate most of the Main Belt \citep{Masiero}.  As discussed in M4, this bias complicates study of the relationship between size and \pv; this relationship is best studied using the full NEOWISE dataset as it is relatively unbiased with respect to \pv.  Figure \ref{fig:carvano_albedo} shows histograms of \pv\ for the various taxonomic types, and Table 1 summarizes their statistical properties.  The \CS\ types are fairly well-grouped but span a wide range of moderate to high \pv; while the \CC\ types tend to have significantly lower \pv, roughly 15\% have \pv$>$0.1, and the \CD\ types have 36\% with \pv$>$0.1.  In M4 we found that spectroscopically classified C types with diameters $>$30 km tended to have \pv$<$0.1.  The fact that so many \CC\ and \CD\ types with diameters $<$30 km have higher albedos suggests either that it is difficult to distinguish these objects with the SDSS colors in the Carvano scheme, that the bias against small objects with low \pv\ is skewing the result, or there is real scatter within the population of \CC\ and \CD\ types.   The bias in the SDSS observations against small, low albedo objects means that measured \pv\ distributions will be skewed higher.  Table 1 gives the statistical properties of the various classes both for all objects and for objects with classifications assigned with probabilities $>50\%$.  As can be seen in the table, removing the objects with low probabilities does not significantly affect the results.

Some asteroids have multiple observations in SDSS, and the different SDSS observations can each result in different classifications. Out of the 24,353 asteroids, 3924 received multiple classifications.  As described in \citet{Carvano}, the most frequent variations occur between classes with adjacent limits, such as \CL\ and \CS, and from \CC\ to \CX.  As can be observed in Figure \ref{fig:carvano_albedo}, the $\cal{L}$S$_{p}$ objects have an albedo distribution that resembles a blend of \CS\ and \CL\ types; similarly, the C$X_{p}$ types' albedos appear to be a cross between \CC\ and \CX\ types.  Of more interest are the objects with classifications that span dissimilar classes, such as \CC\ to \CS or \CC\ to \CQ; however, only a handful of objects in our sample fall into these classes. VNIR spectroscopy of these objects would illuminate what combination of slopes and/or absorption features gives rise to their diverse set of albedos.   

\begin{figure}
\figurenum{3}
\includegraphics[width=6in]{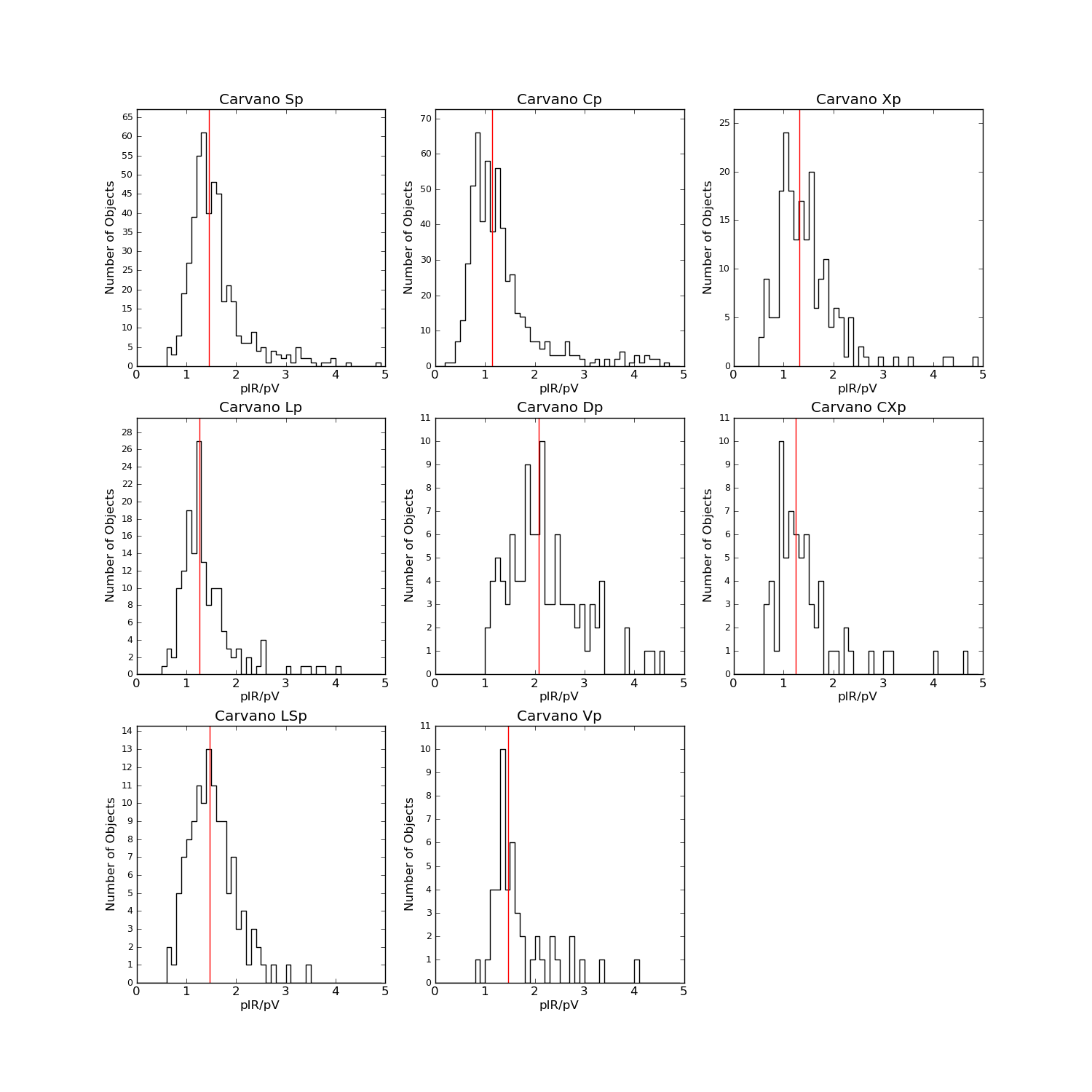}
\caption{\label{fig:carvano_irfactor} NEOWISE-derived ratio \irfactor\ for asteroids observed and classified according to the system of \citet{Carvano}.  Only asteroids for which \irfactor\ could be fitted are included in this plot.  While the \CC\ and \CD\ classes have similar \pv\ values, \irfactor\ is distinctly different. The median \irfactor\ value is shown as a vertical red line.}
\end{figure}

\emph{\CC\ and \CD\ Types.}As discussed in M4, the B, C, D and T types of \citet{Bus02b} and \citet{Tholen84} have similarly low \pv, yet they are distinguished by markedly different \irfactor.  We observe a similar phenomenon here with the \CC\ and \CD\ types.  Figure \ref{fig:carvano_irfactor} shows the distribution of \irfactor\ for those asteroids with sufficient reflected light in $W1$ or $W2$ to compute it.  The \CC\ and \CD\ types have median \pv=$0.064\pm0.001$ and $0.080\pm0.002$, respectively, yet the \CC\ types have \irfactor=$1.147\pm$0.032; \CD\ types have \irfactor$=$2.079$\pm$0.073.  In M4, we hypothesized that the divergent \irfactor\ was caused by fact that the \CC\ types' relatively flat spectra from 0.4 to 1.0 $\mu$m extends out to $3-4\mu$m, while the \CD\ types slope steeply redward in VNIR wavelengths, continuing out to the $W1$ and $W2$ bands.  Figure \ref{fig:scatter_carvano} illustrates the utility of \irfactor\ in distinguishing various ``dark'' asteroid types from one another.  We have made the assumption that $p_{W1}=p_{W2}$ at present and will revisit this assumption in the future.  As discussed above, although the \CC\ and \CD\ asteroids generally have low albedos, they have a tail of higher albedos that are more similar to the bulk of the S-complex objects.  Removing the objects with probabilities lower than $50\%$ did not remove the tail of high-albedo \CC\ and \CD\ objects; this tail of high albedo objects could represent either real scatter within the populations' albedos or difficulty correctly distinguishing the C and D types on the basis of their SDSS photometry alone. 

\emph{X-Complex.}  The distribution of NEOWISE-derived albedos found for \CX\ asteroids closely resembles that observed in the general, undebiased Main Belt population \citep{Masiero}.  We observed a similarly wide range of albedos in M4 for asteroids spectroscopically classified as X-complex.  The wide range of albedos is not surprising, since the Tholen X type (from which the Bus and Bus-DeMeo X types are derived) consists of E, M and P asteroids that are separated by their albedos.  The Carvano \CX\ type is based upon the Bus and Bus-DeMeo X types, neither of which use \pv\ as part of their classification schemes.  

\emph{\CS, \CL, \CA, and \CQ\ Types.} \citet{Bus02b} deemed the S-complex to have sufficient spectroscopic variation to warrant dividing it into six different subclasses.  With the five SDSS bands, the Carvano system does not separate the S-complex. The \CS\ objects span a wide range of albedos with a similar though slightly higher mean value than the \CL\ types.  Their \irfactor\ values largely overlap, though the peak of the distribution shown in Figure \ref{fig:carvano_irfactor} is slightly lower for the \CL\ types.  The \CL\ type is an amalgamation of the Bus K, L and Ld classes, and these are distinguished by the degree of reddening shortward of $i'$ band and flattening longward of $\sim$1 $\mu$m.  The \CA\ types are thought to represent mantle material, and for the 85 objects we observed with NEOWISE, we find that their albedos are very similar to the \CS\ types.  \citet{Bus02b} interpret Q-type asteroids as being the un-space-weathered parent bodies of the ordinary chondrite meteorites; see \citet{Chapman04} for a discussion of the effects of space weathering.  In M4, we found NEOWISE observations of a handful of objects spectroscopically determined to be Q types, but their albedos were identical to the S complex; further, we found no significant differences between the \citet{DeMeo09} S and Sw types, where the Sw type is thought to represent weathered S types.  Here, we find 424 \CQ\ types with cryogenic NEOWISE observations sufficient to determine albedo, and our results reveal that the \CQ\ types have slightly lower \pv\ than the \CS\ types.  All but two of the 424 \CQ\ types are Main Belt asteroids or Mars crossers, and they are distributed throughout the Main Belt as shown in \citet{Carvano}; it would be worthwhile to obtain VNIR spectroscopy of these objects in order to determine whether or not the Q classification is accurate as the Main Belt is generally thought to be bereft of Q types \citep{Binzel10}.  A nearly equal number of asteroids (453) are classified as being intermediate in type between \CS\ and \CQ with extremely similar \pv.

\emph{\CO\ Asteroids}.  \citet{Binzel93} identified (3628) \boznemcova\ as the possible source of the L6 and LL6 ordinary chondrites based on the similarity of its reflectance spectrum to laboratory measurements of these common meteorites. Based on its spectral similarity to (1862) Apollo, they use Apollo's albedo and slope parameter $G$ to compute a diameter of 7 km. From NEOWISE observations of (3628) \boznemcova, we compute a diameter of 7.0$\pm$0.2 km with an albedo of 0.330$\pm$0.043.  Only two asteroids classified as O type spectroscopically according to the taxonomic system of \citet{Bus02b} were observed by NEOWISE, (3628) \boznemcova\ and (169) Zelia, which is classified as an O type by \citet{Lazzaro} using the taxonomic system of \citet{Bus02b}.  Using NEOWISE observations, Zelia has a diameter of 38.6$\pm$0.9 km and an albedo of 0.178$\pm$0.035; its ratio \irfactor=2.08$\pm$0.32.  \citet{Carvano} identified 63 asteroids as O-type candidates, and we have observed 16 with NEOWISE, including \boznemcova.  As shown in Table 1, these objects span a wide range of albedos, from 0.035 to 0.436, with a median value of 0.076; the largest is 12.1$\pm$0.2 km in diameter, and the smallest is 1.75$\pm$0.23 km.  None of these had measurements in $W1$ or $W2$, so \irfactor\ could not be determined.  It would be worth obtaining spectroscopic follow-up of these O-type candidates (asteroids 11753, 20477, 21692, 29540, 32846, 34341, 57736, 104596, 118595, 140949, 164099, 178653, 247450, 261454, 2005 VO101) to see if their spectra confirm their SDSS-based classifications. We note that the \CO\ types with the highest albedos also have the highest probabilities assigned by \citet{Hasselmann}; all of the \CO\ types with \pv$<0.2$ have probabilities lower than 25\%. 

\emph{\CV\ Type.} \citet{Carvano} found a clustering of \CV\ type asteroids near the Vesta family, with some scattered objects throughout the rest of the Main Belt.  We observed 650 \CV\ objects with NEOWISE and find that they consistently have high \pv, tending to be brighter than the \CS\ types.  Since the \CV\ types occupy a distinct region of albedo space compared to the background Main Belt asteroids \citep{Masiero}, it may be possible to use albedo in conjunction with orbital elements to find new candidate members of the Vesta family; this will be explored in a future work.  The orbital elements of these objects are consistent with Vesta family members identified by \citet{Nesvorny10}, with a median semi-major axis of 2.35 AU and standard deviation (SD) of 0.13 AU; median eccentricity of 0.11 with SD$=$0.04; and median inclination of 6.5$^\circ$ with SD$=$2.4$^\circ$. It would be desirable to obtain VNIR spectra of the \CV\ asteroids identified herein to further explore their true natures.  

\begin{figure}
\figurenum{6}
\includegraphics[width=6in]{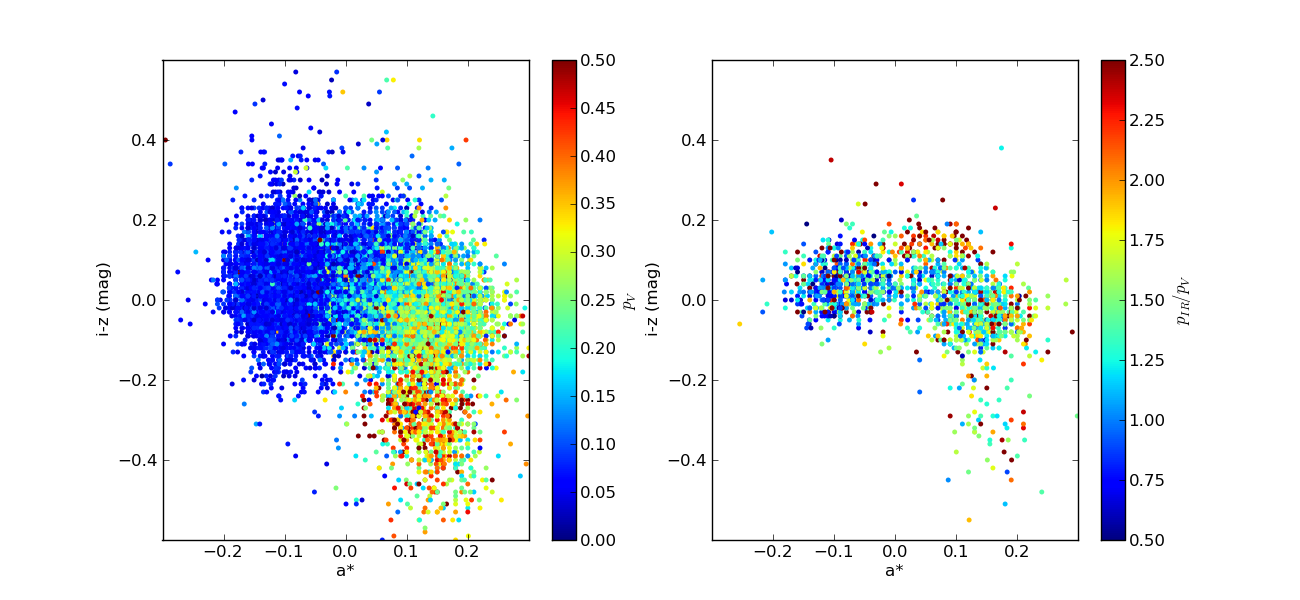}
\caption{\label{fig:a_star} SDSS $a^{*}$ vs. $i-z$ colors for asteroids are shown with \pv\ and \irfactor\ shown as different colored points. }
\end{figure}

Following the work of \citet{Ivezic,Ivezic02}, we have plotted SDSS $a^{*}$ vs. $i-z$ band magnitudes, with albedo shown as colored points (Figure \ref{fig:a_star}); the shading of \CC\ and \CS\ type points appears as expected from these prior works. As noted by \citet{Juric}, asteroids with the bluest $i-z$ colors appear to be associated with the Vesta family, and it can be seen in the figure that these objects tend to have high visible albedos.

\section{Conclusions}
By examining the intersection between the large thermal infrared NEOWISE dataset and the VNIR SDSS observations, we have further explored the relationship between taxonomic classifications and \pv\ for 24,353 asteroids and at 3-4 $\mu$m for 1,819 asteroids.  We find that the Carvano \CS\ and \CV\ types appear to be relatively robust in identifying asteroids with high \pv.  While the \CC\ types are generally significantly darker, we observe in the \CC\ and \CD\ populations high albedo tails that lead us to conclude that albedo cannot be conclusively determined purely from photometric VNIR taxonomy.  We find a number of candidate O-type asteroids among our sample; others have suggested that these are the progenitors of the LL6 ordinary chondrites, as well as Q-type candidates throughout the Main Belt.  The objects with SDSS colors consistent with V-type asteroids typically have very high albedos, and their orbital elements are similar to known Vesta family members, suggesting that these new objects may indeed be family members as well.  As in \citet{Mainzer11d}, we find that \CD\ types, which have a steeply red VNIR spectrum, also have systematically larger \irfactor\ than \CC\ types, despite having similar \pv\ distributions.  These higher \irfactor\ values suggest that the \CD\ types' red spectrum continues through the WISE $W1$ and $W2$ bands at 3-4 $\mu$m.  We caution that since these taxonomic types are determined from objects selected by a visible survey, the population is biased against low albedo objects, and hence the albedo distributions we have determined are similarly biased, particularly at the smallest size scales.  This bias could be mitigated by obtaining a spectrally classified sample that better represents the true population of asteroids, both bright and dark.  Such a sample could be created by choosing a set of objects from the NEOWISE survey, which has been found to be essentially unbiased with respect to \pv\ \citep{Mainzer11e}, and obtaining taxonomic classifications for these objects using a combination of literature values and new observations. While painstaking, such an effort would allow the distinction between observational bias and the physical links between size, albedo, and taxonomic classification to be clarified.   

\section{Acknowledgments}

\acknowledgments{This publication makes use of data products from the \emph{Wide-field Infrared Survey Explorer}, which is a joint project of the University of California, Los Angeles, and the Jet Propulsion Laboratory/California Institute of Technology, funded by the National Aeronautics and Space Administration.  This publication also makes use of data products from NEOWISE, which is a project of the Jet Propulsion Laboratory/California Institute of Technology, funded by the Planetary Science Division of the National Aeronautics and Space Administration. We thank our referee, Dr. Schelte J. Bus, for his comments, which have greatly improved this paper.  We gratefully acknowledge the extraordinary services specific to NEOWISE contributed by the International Astronomical Union's Minor Planet Center, operated by the Harvard-Smithsonian Center for Astrophysics, and the Central Bureau for Astronomical Telegrams, operated by Harvard University.  This research has made use of the NASA/IPAC Infrared Science Archive, which is operated by the Jet Propulsion Laboratory, California Institute of Technology, under contract with the National Aeronautics and Space Administration.}

\clearpage

\begin{deluxetable}{lllllllllllllllll}
\tabletypesize{\tiny}
\rotate
\tablecaption{Median values of \pv\ and \irfactor\ using NEOWISE cryogenic observations of asteroids with taxonomic types derived from SDSS colors according to the method of \citet{Carvano} and \citet{Hasselmann}. The number of objects, median, standard deviation of the mean, standard deviation  (SD), minimum and maximum \pv\ and \irfactor\ are given for all the objects with a particular classification.  Spectral classes with median \irfactor=0.000 did not have enough asteroids with measurements in $W1$ and $W2$ to compute \irfactor. Columns with a ``50" in the heading indicate the statistical properties of only those asteroids that had taxonomic classifications assigned by \citet{Hasselmann} with a probability of 50\% or greater.  }
\tablewidth{0pt}
\tablehead{
\colhead{Class} & \colhead{N} & \colhead{Median} & \colhead{SD} & \colhead{Min} & \colhead{Max} & \colhead{N} & \colhead{Median} & \colhead{SD} & \colhead{Min} & \colhead{Max} & \colhead{N} & \colhead{Median} & \colhead{SD} & \colhead{N 50} & \colhead{Median 50} & \colhead{SD} \\

\colhead{} & \colhead{\pv} & \colhead{\pv} & \colhead{} & \colhead{} & \colhead{} & \colhead{\irfactor} & \colhead{\irfactor} & \colhead{SD} & \colhead{Min} & \colhead{Max} & \colhead{\pv50} & \colhead{\pv50} & \colhead{50} & \colhead{\irfactor} & \colhead{\irfactor} & \colhead{50}

}
\startdata
Sp & 4880 & $ 0.262 \pm 0.001$ & 0.084  & 0.034 & 0.779 & 474 & $ 1.452 \pm 0.028$ & 0.603  & 0.607 &  4.945 & 2725 & $ 0.267 \pm 0.002 $ & 0.080 & 362 & $ 1.449 \pm 0.030 $ & 0.570 \\
Cp & 9779 & $ 0.064 \pm 0.001$ & 0.055  & 0.013 & 1.000 & 566 & $ 1.147 \pm 0.032$ & 0.765  & 0.260 &  4.971 & 3069 & $ 0.067 \pm  0.001 $ & 0.054 & 374 & $ 1.095 \pm 0.035 $ & 0.680 \\
Xp & 1773 & $ 0.106 \pm 0.003$ & 0.118  & 0.020 & 1.000 & 201 & $ 1.318 \pm 0.044$ & 0.625  & 0.507 &  4.801 & 502 & $ 0.109 \pm  0.006 $ & 0.127 & 138 & $ 1.322 \pm 0.056 $ & 0.658 \\
Lp & 1838 & $ 0.202 \pm 0.002$ & 0.085  & 0.030 & 0.638 & 155 & $ 1.263 \pm 0.047$ & 0.580  & 0.577 &  4.036 & 768 & $ 0.204 \pm  0.003 $ & 0.087 & 109 & $ 1.252 \pm 0.050 $ & 0.518 \\
Dp & 984 & $ 0.080 \pm 0.002$ & 0.073  & 0.014 & 0.516 & 104 & $ 2.079 \pm 0.073$ & 0.744  & 1.034 &  4.543 & 235 & $ 0.073 \pm  0.003 $ & 0.041 & 67 & $ 2.165 \pm 0.094 $ & 0.766 \\
Ap & 85 & $ 0.248 \pm 0.009$ & 0.087  & 0.081 & 0.488 & 9 & $ 2.526 \pm 0.212$ & 0.637  & 1.307 &  3.166 & 20 & $ 0.273 \pm  0.018 $ & 0.082 & 4 & $ 2.187 \pm 0.262 $ & 0.523 \\
Qp & 424 & $ 0.253 \pm 0.005$ & 0.110  & 0.038 & 0.613 & 14 & $ 1.456 \pm 0.106$ & 0.398  & 0.939 &  2.144 & 118 & $ 0.295 \pm  0.009 $ & 0.093 & 7 & $ 1.100 \pm 0.172 $ & 0.455 \\
Op & 16 & $ 0.076 \pm 0.039$ & 0.157  & 0.035 & 0.436 & 0 & $    0.000 \pm    0.000$ & 0.000  & 0.000 &  0.000 & 0 & $ 0.000 \pm  0.000 $ & 0.000 & 0 & $ 0.000 \pm 0.000 $ & 0.000  \\
Vp & 650 & $ 0.343 \pm 0.004$ & 0.105  & 0.047 & 0.771 & 47 & $ 1.470 \pm 0.092$ & 0.631  & 0.886 &  4.046 & 288 & $ 0.352 \pm  0.006 $ & 0.100 & 29 & $ 1.560 \pm 0.099 $ & 0.532  \\
CSp & 1 & $ 0.130 \pm 0.000$ & 0.000  & 0.130 & 0.130 & 0 & $ 0.000 \pm 0.000$ & 0.000  & 0.000 &  0.000 & 0 & $ 0.000 \pm  0.000 $ & 0.000 & 0 & $ 0.000 \pm 0.000 $ & 0.000  \\
XSp & 36 & $ 0.203 \pm 0.017$ & 0.100  & 0.044 & 0.507 & 2 & $ 0.885 \pm 0.040$ & 0.056  & 0.829 &  0.941 & 1 & $ 0.507 \pm  0.000 $ & 0.000 & 1 & $ 0.829 \pm 0.000 $ & 0.000  \\
LSp & 1488 & $ 0.240 \pm 0.002$ & 0.083  & 0.042 & 0.623 & 124 & $ 1.476 \pm 0.043$ & 0.479  & 0.673 &  3.472 & 396 & $ 0.247 \pm  0.004 $ & 0.083 & 79 & $ 1.449 \pm 0.054 $ & 0.480  \\
SQp & 453 & $ 0.252 \pm 0.004$ & 0.092  & 0.062 & 0.617 & 17 & $ 1.561 \pm 0.117$ & 0.484  & 1.069 &  3.085 & 93 & $ 0.262 \pm  0.009 $ & 0.084 & 8 & $ 1.444 \pm 0.098 $ & 0.276  \\
SAp & 31 & $ 0.273 \pm 0.016$ & 0.091  & 0.135 & 0.451 & 3 & $ 1.672 \pm 0.125$ & 0.216  & 1.299 &  1.811 & 0 & $    0.000 \pm     0.000 $ & 0.000 & 0 & $    0.000 \pm    0.000 $ & 0.000  \\
SVp & 44 & $ 0.316 \pm 0.015$ & 0.097  & 0.111 & 0.549 & 2 & $ 1.874 \pm 0.407$ & 0.575  & 1.299 &  2.450 & 0 & $    0.000 \pm     0.000 $ & 0.000 & 0 & $    0.000 \pm    0.000 $ & 0.000  \\
CXp & 1144 & $ 0.067 \pm 0.003$ & 0.087  & 0.020 & 0.999 & 67 & $ 1.243 \pm 0.104$ & 0.851  & 0.643 &  5.031 & 104 & $ 0.069 \pm  0.011 $ & 0.110 & 20 & $ 1.122 \pm 0.062 $ & 0.276  \\
CDp & 22 & $ 0.064 \pm 0.012$ & 0.057  & 0.030 & 0.240 & 0 & $ 0.000 \pm 0.000$ & 0.000  & 0.000 &  0.000 & 0 & $ 0.000 \pm  0.000 $ & 0.000 & 0 & $ 0.000 \pm 0.000 $ & 0.000  \\
CQp & 6 & $ 0.139 \pm 0.017$ & 0.042  & 0.083 & 0.192 & 0 & $ 0.000 \pm 0.000$ & 0.000  & 0.000 &  0.000 & 0 & $ 0.000 \pm  0.000 $ & 0.000 & 0 & $ 0.000 \pm 0.000 $ & 0.000  \\
XLp & 170 & $ 0.159 \pm 0.005$ & 0.060  & 0.027 & 0.452 & 10 & $ 1.116 \pm 0.122$ & 0.387  & 0.491 &  2.015 & 26 & $ 0.160 \pm  0.016 $ & 0.081 & 5 & $ 1.059 \pm 0.099 $ & 0.221  \\
XDp & 209 & $ 0.076 \pm 0.006$ & 0.091  & 0.018 & 0.715 & 11 & $ 1.628 \pm 0.172$ & 0.569  & 0.996 &  2.689 & 0 & $    0.000 \pm     0.000 $ & 0.000 & 0 & $    0.000 \pm    0.000 $ & 0.000  \\
DLp & 170 & $ 0.178 \pm 0.007$ & 0.091  & 0.044 & 0.531 & 9 & $ 1.610 \pm 0.148$ & 0.445  & 1.062 &  2.456 & 2 & $ 0.214 \pm  0.085 $ & 0.120 & 0 & $    0.000 \pm    0.000 $ & 0.000  \\
QVp & 40 & $ 0.352 \pm 0.015$ & 0.097  & 0.088 & 0.575 & 1 & $ 1.488 \pm 0.000$ & 0.000  & 1.488 &  1.488 & 0 & $    0.000 \pm     0.000 $ & 0.000 & 0 & $    0.000 \pm    0.000 $ & 0.000  \\
QLp & 5 & $ 0.249 \pm 0.031$ & 0.069  & 0.145 & 0.361 & 0 & $ 0.000 \pm 0.000$ & 0.000  & 0.000 &  0.000 & 0 & $ 0.000 \pm  0.000 $ & 0.000 & 0 & $ 0.000 \pm 0.000 $ & 0.000  \\
\enddata
\end{deluxetable}

\begin{figure}
\figurenum{4}
\includegraphics[width=6in]{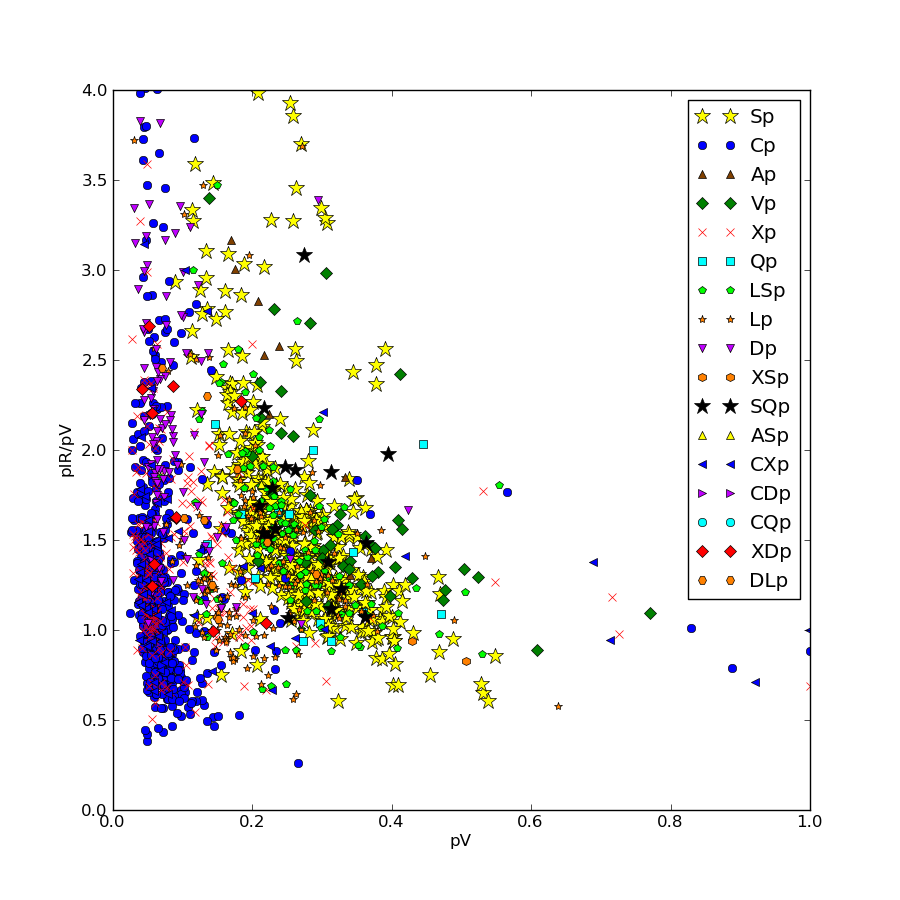}
\caption{\label{fig:scatter_carvano} NEOWISE-derived ratio $p_{IR}/p_{V}$ vs. $p_{V}$ for asteroids observed and classified according to the Carvano taxonomic classification scheme.  Only the $\sim$1800 asteroids for which $p_{IR}/p_{V}$ could be fitted are included in this plot.}
\end{figure}

\begin{figure}
\figurenum{5}
\includegraphics[width=3in]{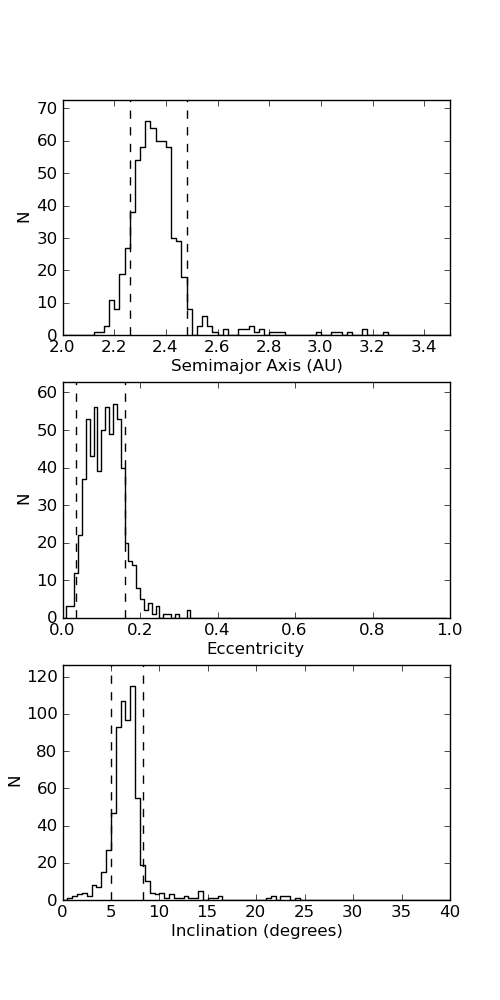}
\caption{\label{fig:scatter_carvano} Orbital elements of the 650 asteroids classified as \CV; the typical ranges for the Vesta family are shown as dashed vertical lines \citep{Nesvorny10}.}
\end{figure}

\end{document}